\documentclass[aps,prl,twocolumn, superscriptaddress, amsmath]{revtex4-1}

\usepackage{natbib}
\usepackage{graphicx}
\usepackage{bm}
\usepackage{color,soul}
\usepackage{hyperref}
\usepackage{float}
\usepackage{csquotes}
\usepackage{color,soul}
\usepackage{hyperref}
\hypersetup{colorlinks=true, urlcolor=blue, citecolor=blue}
\usepackage{longtable}
\usepackage{tabularx}

\newcommand{\cgx}{{\rm CsGeX$_3$}}
\newcommand{\cgb}{{\rm CsGeBr$_3$}}
\newcommand{\cgi}{{\rm CsGeI$_3$}}
\newcommand{\bfo}{{\rm BiFeO$_3$}}

\begin{document}

\title{Theory of ultrathin ferroelectrics: the case of CsGeBr$_3$ }

\author{Ravi Kashikar}
\email{ravik@usf.edu}
\author{Arlies Valdespino}
\author{Charlton Ogg}
\author{Edvin Uppgard}
\author{S. Lisenkov}
\author{I. Ponomareva}
\email{iponomar@usf.edu}
\affiliation{Department of Physics, University of South Florida, Tampa, Florida 33620, USA}

\date{\today}

\begin{abstract}
Ferroelectricity has recently been demonstrated in germanium-based inorganic halide perovskites. We use atomistic first-principles-based simulations to study ultra-thin \cgb\ films with thicknesses of  4-18 nm and develop a theory for ferroelectric ultrathin films.  
The theory introduces (i) a local order parameter, the local polarization, which allows the identification of phase transitions into both monodomain and polydomain phases, and (ii) a dipole pattern classifier, which allows efficient and reliable identification of unique dipole patterns. Application of the theory to both halides \cgb\ and \cgi\, as well as oxide \bfo\ ultrathin ferroelectrics, which undergo paraelectric cubic to ferroelectric rhombohedral phase transition in bulk, reveal two distinct scenarios for ultrathin films. In the first one, the films transition into a monodomain phase, which is allowed below a critical value of the residual depolarizing field. Above this critical value, the second scenario occurs, and the film undergoes a phase transition into a nanodomain phase. The two scenarios are associated with the opposite response of Curie temperature to thickness reduction. As the film's thickness decreases, the transition temperature into the monodomain phase increases while the transition temperature into the nanodomain phase decreases. The surface effects are responsible for the Curie temperature enhancement, while the stripe domain pattern is the origin of the transition temperature suppression.  Application of dipole pattern classifier reveals a rich variety of nanodomain phases in halide films: nano-stripes, labyrinths, zig-zags, pillars, and lego-domains.  Our work could lead to both a deeper understanding of nanoscale ferroelectrics and discoveries of unusual nanoscale dipole patterns.

\end{abstract}
\maketitle

Ferroelectrics in thin film form have proven to be excellent functional material for memory devices,  sensor technologies, microelectromechanical systems (MEMS), and energy harvesting devices \cite{Scott-Apll, Scott, martin2016thin, science_Rabe}. In the thin film form, ferroelectrics could exhibit high Curie temperature(T$_C$), large remnant polarization, small coercive fields, and unique structural phases as compared to their bulk phase \cite{RABE2005122}.  For example, Curie temperature for thin BaTiO$_3$ films is nearly 600~K, which is a 200~K enhancement with respect to bulk,  while the remnant polarization could be enhanced by as much as 250\%  \cite{Thin_BTO}. The unique characteristics of thin films are believed to arise from the interplay of growth directions, deposition conditions, the residual depolarizing field, thickness, and epitaxial strain \cite{Growth, strain, mehta1973depolarization}. These factors not only differentiate the films from their bulk counterparts but also give rise to unconventional dipole patterns and nanodomains, as well as topological structures\cite{dipole-1,dipole-2,Polar_Topo}. A few fascinating examples are stripe domains in PbTiO$_3$ and BaTiO$_3$ \cite{Streifer,PBT_stripe, BTO_stripe}, bubble domains in Pb(Zr, Ti)O$_3$\cite{Bubble}, and labyrinths in ultrathin BiFeO$_3$ films \cite{ Nahas2020}. Recently, freestanding ferroelectric membranes have been realized, opening new opportunities for property manipulation and engineering \cite{Free_Mem-1,Xu2020}. Interestingly, most investigations into these features have focused on oxide ferroelectrics. However, oxide ferroelectrics, in general, pose challenges for flexible device applications due to their inherent hardness originating from strong ionic and covalent bonds\cite{gao2015flexible}.

The halide perovskites' soft nature and semiconducting properties are attractive features for flexible device applications\cite{halide-rev}. Recently, various organic-inorganic halide perovskites MPSnBr$_3$, TMCM-(Mn, Cd)Cl$_3$,  have been identified as ferroelectric materials, with spontaneous polarization ranging 3-8 $\mu$C/cm$^2$, which is significantly lower than that of oxide  ferroelectrics\cite{shahrokhi2020emergence, zheng2023emerging}. The potential to overcome this challenge is offered by the recent experimental confirmation of ferroelectricity in inorganic halide perovskites with a spontaneous polarization of 15-20 $\mu$C/cm$^2$\cite{CGX_Ferro}. 
These findings provide new hope for alternative materials for both device applications and scientific discoveries. Furthermore, these materials are already recognized as semiconductors and are actively investigated from an optoelectronic perspective\cite{CGI_solarcell}, which could open the way to additional multifunctionality. At present, there are no studies available on the effect of scaling down of \cgx\ (X = Cl, Br, I) halide perovskites. Consequently, it is not known how the ferroelectric properties, such as ferroelectric phases, transition temperatures, and dipole patterns, respond to thickness reduction. Do halide perovskites develop monodomain or polydomain phases at the nanoscale, and what could be the nanodomain patterns? Can they sustain residual depolarizing fields, and if yes, up to what strength? One can hypothesize that by manipulating the residual depolarizing field in semiconducting halide perovskites, we can potentially enhance Rashba effects, opening new opportunities to explore fundamental science and potential applications \cite{Rashba_appl}. Likewise, the effects that epitaxial strain or surface charge screening plays in the ferroelectricity of such films are presently unknown. Therefore, in this study, we take advantage of recently developed first-principles-based effective Hamiltonian \cite{RAVI_CGB} to (i) investigate ferroelectricity in \cgx\ (X = Br, I) ultrathin films;  (ii)  develop a theory of ferroelectric ultrathin films, which establishes and elucidates the effects of scaling down; (iii) reveal that \cgb\ films can sustain extraordinary high residual depolarizing fields (up to 2.8 GV/m), promising for potential applications in spintronics; (iv) report a rich variety of dipole patterns, which can be induced in such films.

To achieve our goals, we use the first principle-based effective Hamiltonian, which has been developed for \cgx\ (X = Br, I)\cite{RAVI_CGB, unpublishedkeyB}. The degrees of freedom for the Hamiltonian are local modes, which are proportional to the local dipole moment of the unitcell, and homogeneous and inhomogeneous strain\cite{Effect_H0, Effective_H1}. This approach has proven highly effective in capturing intriguing features in bulk and in low-dimensional oxide ferroelectrics\cite{BFO,BST,BZT,PTO_phase,PZo_phase,PZT}. Recently, it has been used to study phase transitions in \cgb\ under different mechanical loads \cite{Josh_prb}.
To simulate the \cgb\ thin films, we considered the supercell of 24$\times$24$\times$N$_z$ unitcells of \cgb, with periodic boundary conditions applied along the in-plane directions. The thickness along the growth direction is modeled by N$_z$=8, 12, 16, 20, 24, and 32, which simulates the range from 4 nm to 18 nm. The z cartesian axis points along the film's growth direction.
The total energy for the film's supercell is given by \cite{Effective_H1}
\begin{widetext}
    \begin{equation}
  E_{tot} = E_{self}(\{\mathbf{u}_i\})+E_{dpl}(\{\mathbf{u_i}\})+E_{short}(\{\textbf{u}_i\})+E_{elas}(\{\eta_l\})+E_{int}(\{\textbf{u}_i\}, \{\eta_l\})+\beta \sum_i \langle E_{dep}^{max} \rangle Z^{*} u_i  
\end{equation}
\end{widetext}

where $u_i$ is a local mode of the unitcell $i$, and $\eta_l$ contains both inhomogeneous and homogeneous  strain variables. 
The terms on the right-hand side are the local mode self-energy (harmonic and anharmonic contributions), long-range dipole-dipole interactions, a short-range interaction between
local modes, elastic energy, the interaction between
the local modes and strains, depolarizing field screening term, respectively. The parameter $\beta$ controls the strength of maximum depolarizing field $\langle$E$_{dep}^{max}\rangle$ developed along the z direction is screened by the last term in Eq.~(1), where  $\beta$ parameter controls the strength of the surface charge screening. For example, $\beta$=0 corresponds to ideal open circuit conditions and $\beta$=1 is for ideal short circuit conditions.  We varied $\beta$ from 0.85 to 0.99 in the present case.  We use r$^2$SCAN parameters for the effective Hamiltonian reported in the Ref.\cite{RAVI_CGB, unpublishedkeyB}. The parametrization predicts the Curie temperature of bulk \cgb\ to be 360~K, which underestimates the experimental one, 511~K \cite{Thiele}. The Hamiltonian is used in the framework of classical Molecular Dynamics (MD) simulations, with Newton's equations of motion numerically solved using the predictor-corrector algorithm with an integration step of 1~fs.  The Evans-Hoover thermostat is applied to maintain a constant temperature\cite{rapaport2004art}. To obtain equilibrium phases, the films are annealed from 600~K to 10~K with a step of 10~K. For each temperature, 300,000 MD steps are used.

\begin{figure*}
\centering
\includegraphics[width=1\textwidth]{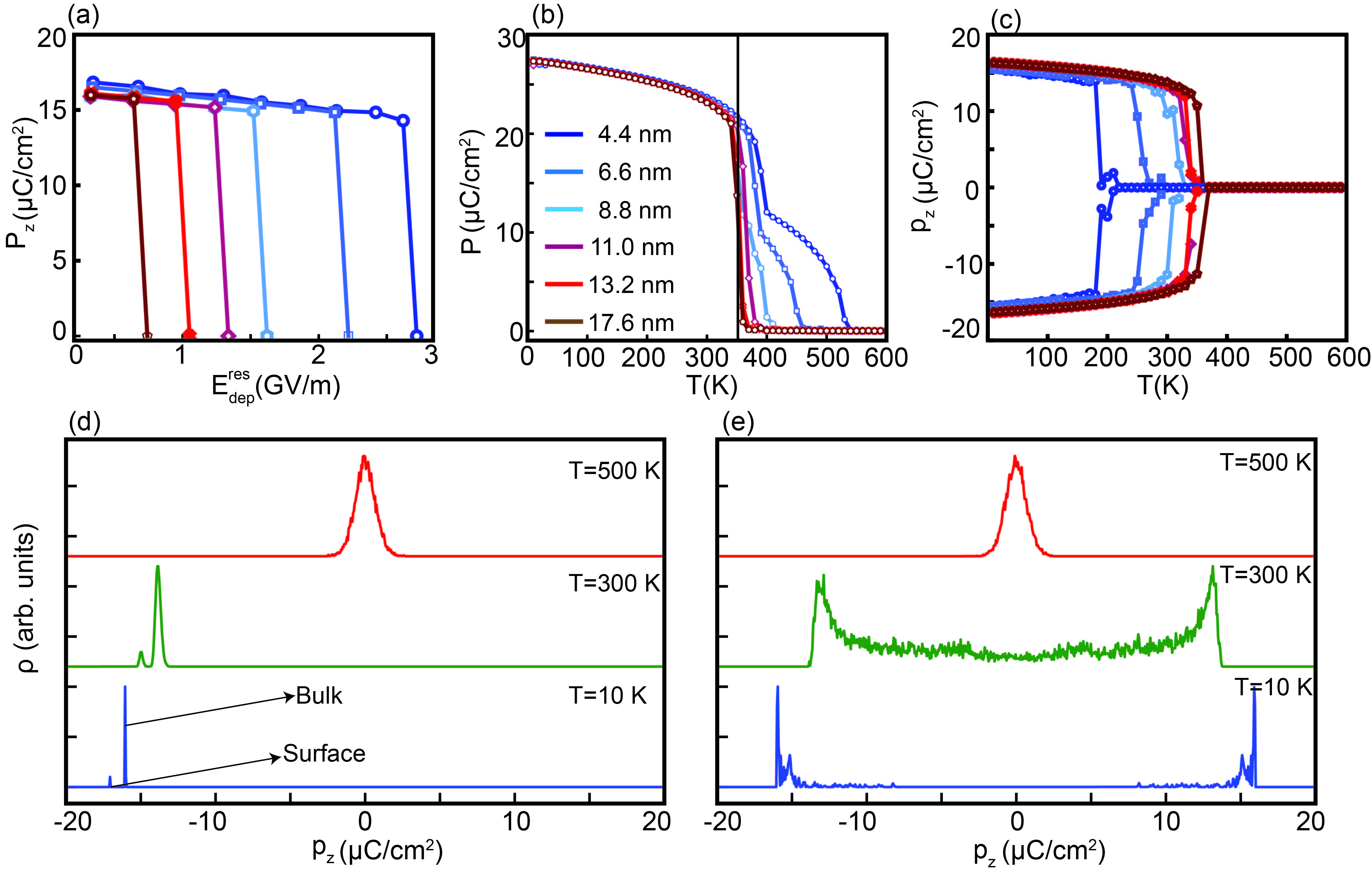}
\caption{ (a) The $P_z$  component of the polarization as a function of the residual depolarizing field at 10 K. (b) Polarization as a function of temperature for films with E$^{res}_{dep}$=0.65 GV/m. The vertical line indicates the T$_C$ for bulk.  (c) The temperature evolution of the local polarization in the nanodomain phase. Distribution of z component of the local polarization at various temperatures for monodomain (d) and nanodomain (e) regimes. }
\label{fig1}
\end{figure*}

We begin by investigating how ferroelectric properties, such as spontaneous polarization and Curie temperature, evolve with the film's thickness. For each film, we ran an annealing simulation for different values of surface charge compensation $\beta$, which is related to the depolarizing field as $E_{dep}=\frac{4\pi P_z(1-\beta)}{\epsilon_0 \epsilon_{\infty}}$,  where P$_z$ is the out-of-plane polarization and $\epsilon_{\infty}$ is the optical dielectric constant of the material. It is well established for oxide perovskites that as the quality of surface charge compensation deteriorates, the residual depolarizing field increases and eventually becomes so strong that the film is not able to maintain the out-of-plane polarization component and undergoes a transition into a nanodomain phase\cite{PBT_stripe,BTO_stripe}. We find a similar trend for \cgb\ ultrathin films. Figure~\ref{fig1}(a) shows the dependence of spontaneous polarization along the growth directions on the residual depolarizing field for films of different thicknesses in the regime of partial surface charge compensation. As the surface charge screening deteriorates, the residual depolarizing field increases while polarization decreases slightly. Once the critical value of the E$_{dep}^{crit}$ is reached, the film transitions into the nanodomain phase, associated with zero out-of-plane polarization component. We can see that the critical E$_{dep}^{crit}$ depends strongly on the thickness of the film and, surprisingly, is the smallest for the thickest film. In other words, the thinnest films are able to sustain larger depolarizing fields in the monodomain phase. Strikingly, a 4.4 nm (8 u.c.) thick film can sustain a 2.8 GV/m of residual depolarizing field. Such a huge built-in field is extremely promising for photovoltaic applications, where it can be used to separate carriers and for spintronics applications. Indeed, \cgx\ (X = Br, I) was found to possess spin splitting in the band structure in a weakly relativistic regime\cite{unpublished_p}. The large built-in electric field could enhance these effects due to Rashba interactions.

Let us now focus on the films in the monodomain phase, that is, before the critical value of the residual depolarizing field is reached. The dependence of polarization on the temperature for films in such a regime is presented in Fig.~\ref{fig1}(b). The data reveals that the T$_C$ is strongly enhanced in the thinnest films. The enhancement is associated with the formation of a tetragonal phase above the bulk T$_C$ with an out-of-plane polarization component. To reveal the origin of such an enhancement, we turn to Fig.~\ref{fig1}(d), which shows the probability density functions for the z-component of the local polarizations, that is the polarization of a unitcell, for N$_z$=24 u.c. thick film computed at different temperatures. At the lowest temperature, there are two well-defined peaks. The largest one coincides with the one in bulk \cgb\ \cite{RAVI_CGB}, while the smallest one is at the larger value of $|p_z|$.  This peak was found to be associated with the dipoles at the surface, suggesting that the surface effect causes the enhancement of the local polarization near the surface. Such polarization enhancement at the surface gives origin to the total polarization enhancement in the thinnest film, where the ratio of the surface to bulk dipoles is the largest. 

Let us now focus on the nanodomain phase of the films, that is, the one that occurs beyond E$_{dep}^{crit}$. Now, the polarization is not a suitable order parameter as its out-of-plane component is zero. To overcome this issue, we again turn to the probability density functions for the $p_z$ component of the local dipoles, which are given in Fig.~\ref{fig1}(e). We can see that they feature two peaks at the lowest temperatures, located at $\pm p_z$. The peaks are associated with dipoles in nanodomains with opposite out-of-plane polarization components. As temperature increases, the peaks merge into a single broad peak centered at $p_z=0$. We now use the location of the maxima of the peaks $<p_z>$ as a new order parameter, the local polarization.  Its temperature evolution is given in Fig.~\ref{fig1}(c) and clearly reveals the transition into the polydomain phase. To test the validity of our proposed local order parameter $<\mathbf p>$, we compare its temperature evolution with polarization for the films in the monodomain phase (See Fig.~S1 of the supplementary material) and find no discrepancies in either the Curie temperature or polarization. Interestingly, the multiplicity of the local order parameter reveals whether the transition is into a polydomain or a monodomain phase. For example, in Fig.~\ref{fig1}(e), there are two values of the local order parameter associated with a given temperature, revealing a transition into a polydomain phase. If, however, the order parameter is single-valued, then the transition is associated with a monodomain phase.

\begin{figure*}
\centering
\includegraphics[width=1\textwidth]{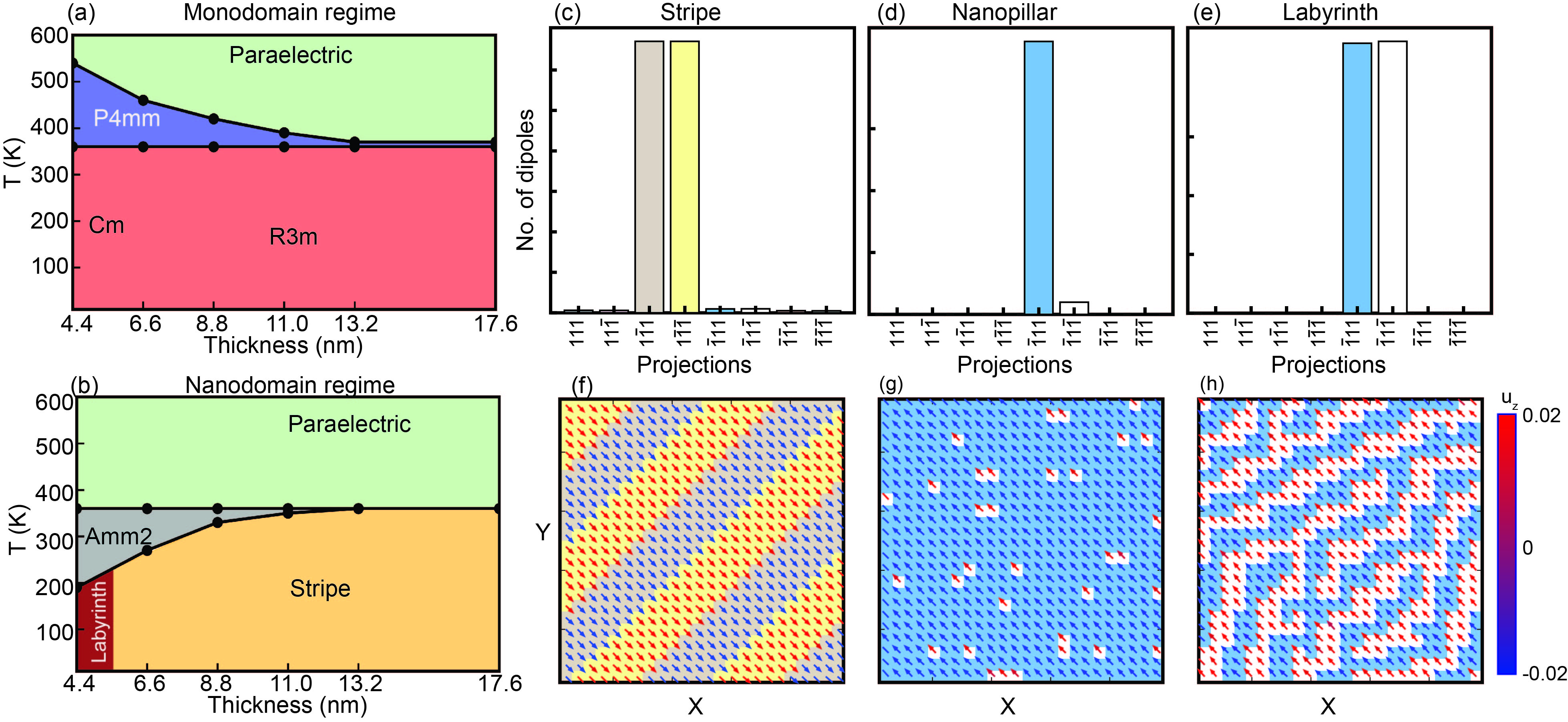}
\caption{  Temperature-thickness phase diagram for \cgb\, thin films in the monodomain (a) and nanodomain (b) regime. For the monodomain regime E$_{dep}^{res}$=0.65~GV/m.   (c)-(e) Class distribution of electric dipoles in the supercell and corresponding dipole patterns  (f)-(h). Panels (f) and (e)-(h) correspond to $N_z=$24 and $N_z$=8, respectively. }
\label{fig2}
\end{figure*}

We can now apply this analysis to study phase transitions in \cgb\ films. Figure~\ref{fig2}(a)-(b) presents phase diagrams for the films in the mono- and nanodomain regimes, respectively.  For the films in the monodomain regime, we find that Curie temperature decreases with thickness, and the film thickness below 11 nm undergoes two phase transitions: paraelectric cubic to ferroelectric P4mm and to R3m phase. The transition temperature associated with the transition into R3m-phase does not change with thickness and coincides with the transition temperature of bulk \cgb\. The reason for the increase in T$_C$ with the decrease in the thickness has already been discussed.   For the films in the nanodomain regime, we find two phase transitions. For films 11 nm and below, the first one is to the ferroelectric Amm2 phase, the second one is into the nanodomain phase with the rhombohedral (R) phase inside each domain. The T$_C$ does not change with thickness and coincides with the T$_C$ of bulk \cgb. The transition temperature for the transition into the nanodomain phase increases as a function of the film's thickness. However, before this finding can be explained, we need to examine the nanodomain phases. The dipole pattern associated with the nanodomain phase in N$_z$=24 thick films at 10 K is given in Fig.~\ref{fig2}(f) and shows the nanostripe domain pattern. Similar nanostripes have been found in oxide ferroelectrics\cite{180_stripe}. As the thickness of the film increases, the stripe width increases similar to the BiFeO$_3$ films\cite{BFO_Kittel} and as expected from the Kittel law. For the thinnest film of N$_z$=8, we find the disconnected labyrinth pattern (Fig.~\ref{fig2}(h)) as well as a nanopillar pattern (Fig.~\ref{fig2}(g)). Previously, labyrinth patterns have been found for PZT  ultra-thin films\cite{Nahas2020}. The nanopillar patterns are found at the boundary between the monodomain and nanodomain phases, that is close to E$_{dep}^{crit}$. We can now understand why the transition temperature into the polydomain phase increases with the film thickness. As the thickness of the film increases, the nanostripes become wider, following the Kittel's law. The increase in the stripe width can be considered as moving in the Brillouin zone from X to $\Gamma$ point. Indeed, the X point is associated with the antipolar phase, that is, nanostripes of 1 unitcell width, while the $\Gamma$ point is associated with an infinitely wide single stripe (monodomain phase).  
The phonon dispersion curves presented in Ref.\cite{RAVI_CGB} for \cgb\ reveal that the phonon frequency decreases as we move from X to $\Gamma$ point, which could be interpreted as the stability of the phase increases as domains become wider. As the nanostripes become wider as the film gets thicker, this explains why the nanodomain phase can be stabilized at higher temperatures for the thicker films. 
\begin{figure*}
\centering
\includegraphics[width=1\textwidth]{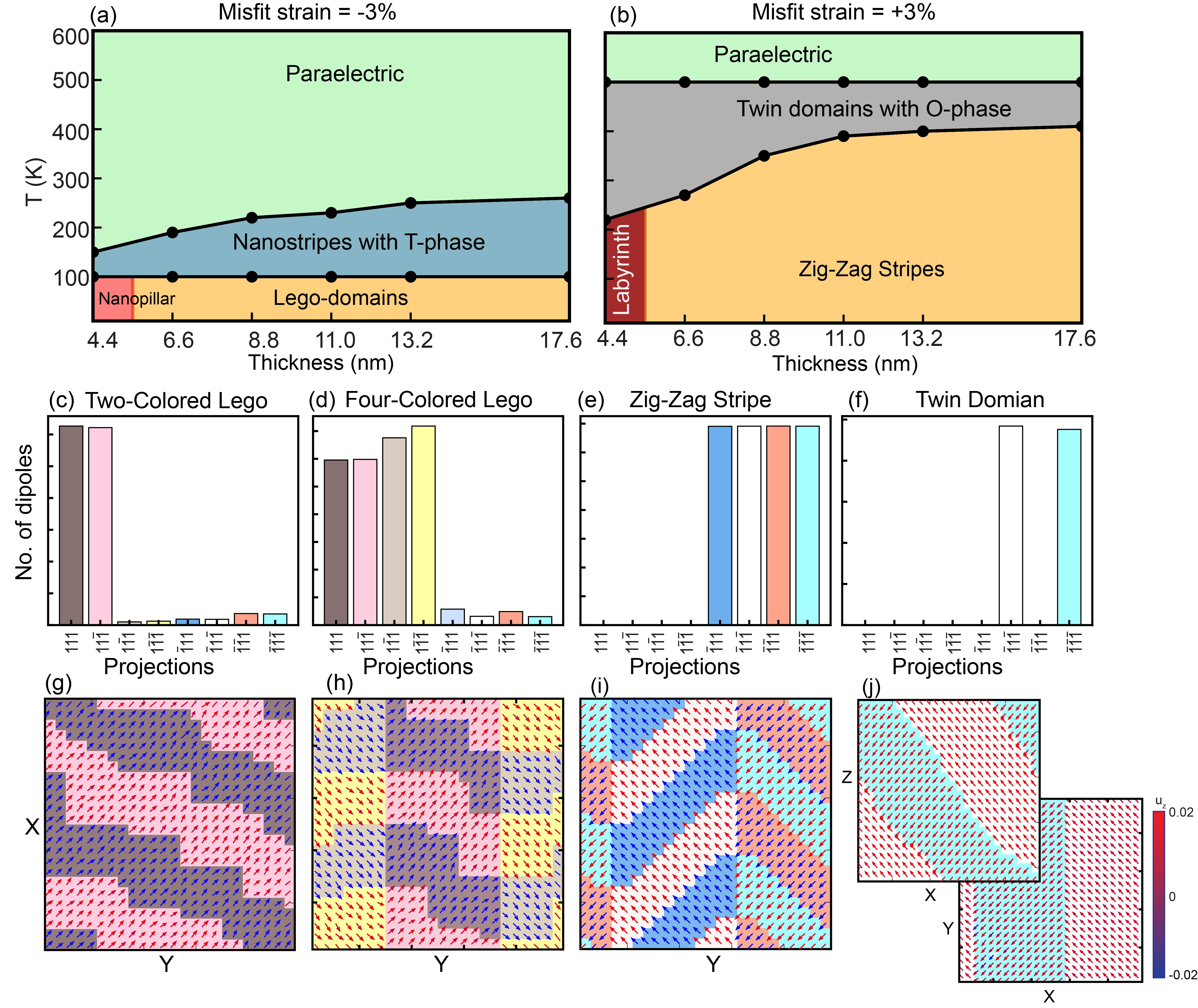}
\caption{(a)-(b) Temperature-thickness phase diagram  for stressed thin films for SC conditions
(c)-(f) Class distribution of dipole orientations and corresponding domain structures in (g)-(I). Panels (g)-(i) and (j) correspond to $N_z=$24 and $N_z$=20, respectively. }
\label{fig3}
\end{figure*}

Our next goal is to determine whether the nanodomain patterns can be controlled through the epitaxial strain. For that, we repeat simulations for compressive and tensile strains of $\pm$3\%. Here, we find a variety of nanodomain patterns. In order to characterize them, we propose the following approach. Each nanodomain exhibits a rhombohedral-like phase with local polarization aligned along one of the eight polar directions $\langle$111$\rangle$. Let us associate these eight directions with eight classes and make the distribution function for the dipoles among these classes. The number of classes and their population tell us about the distinct types of domains we have in a pattern. For example, for nanostripes, we find only two classes associated with   $[\Bar{1}11]$,  $[\Bar{1}1\Bar{1}]$ polarization nanodomains with equal distribution of the dipoles between the classes (see Fig.~\ref{fig2}(c)). The same is true for the labyrinth pattern (see Fig.~\ref{fig2}(e)). However, for the nanopillar pattern, we find that the two classes are unevenly populated, clearly indicating that this is a qualitatively different nanodomain pattern. 

The phase diagrams associated with the nanodomain regime for the strained films are given in Fig.~\ref{fig3}(a)-(b). For the compressive strain, we find two phase transitions into the nanodomain phase. The one at $T_C$ is associated with the formation of nanostripes with tetragonal phases inside different domains. As the temperature is lowered further the nanostripes undergo another phase transition associated with a rhombohedral phase inside each stripe. The nanodomains now are reminiscent of lego blocks (termed as lego-domains) and could be of two colors or multi-colored (see Figs.~\ref{fig3}(g)-(h) ).  These nontrivial patterns are formed as a compromise between (i) the intrinsic tendency of \cgb to have a rhombohedral phase,  (ii) compressive strain dictating to annihilate the in-plane polarization component, and (iii) the depolarizing field working to annihilate out-of-plane polarization component. The associated class distributions for such lego-domain patterns are given in Fig.~\ref{fig3}(c)-(d).

For the films under tensile strain, we find nanodomain patterns even below the critical depolarizing field. Such twin domain patterns are visualized in Fig.~\ref{fig3}(j) and allow the enhance the local in-plane polarization component as required by the tensile strain. The phase diagram for such films is shown in Fig.~\ref{fig3}(b). The nanodomain pattern is now a zig-zag one with four classes (see Fig.~\ref{fig3}(e)). This is the consequence of the twin domains splitting to create two more classes with opposite z-component of polarization (compare class distribution between Fig.~\ref{fig3}(e) and \ref{fig3}(f)). Interestingly, the trends in transition temperatures established for the stress-free films are present for the strained films as well.

We have carried out the same simulations for another member of the halide family, \cgi\, along with the oxide perovskite \bfo\, except for the strained simulations.  To simulate  BiFeO$_3$, we used effective Hamiltonian parametrization of Ref. \cite{BFO_params}.  To \bfo\, we use parameters from Ref.\cite{BFO_params}. We find the same qualitative results, indicating that our findings are applicable to ultrathin ferroelectric films that undergo cubic to rhombohedral phase transitions in bulk. The data from such calculations are given in Supplementary Materials.

In summary, we used atomistic first-principles-based simulations to investigate ultrathin film of ferroelectric \cgb\, with thicknesses 4-18 nm. We found two scenarios for phase transitions in such films, which are controlled by the quality of surface charge screening. When the surface charge is well compensated, the films undergo a phase transition into a monodomain phase. As the quality of surface charge compensation deteriorates, the residual depolarizing field increases, and at some critical value of the field, the second scenario occurs where the films undergo a phase transition into a nanodomain phase. The \cgb\, films can accommodate a very large residual depolarizing field of up to 2.8~GV/m before the second scenario takes place. Interestingly, the field decreases as the thickness of the film increases.   In the nanodomain regime, the films develop nanostripe, nanopillar, or labyrinth dipole patterns. Transition temperatures for the two scenarios have opposite behavior upon scaling down. For the monodomain regime, the Curie point increases as the film's thickness decreases. However, the temperature associated with the transition into the nanodomain phase decreases as the film's thickness decreases. Epitaxially strained films exhibit a wide variety of dipole patterns, including zig-zag nanostripes and lego-domains. Interestingly, the scaling-down trends revealed for \cgb\ films hold for other materials with the same type of phase transition in bulk ( \cgi\, and oxide perovskite \bfo), strongly suggesting that they are universal. We believe that our study lays the foundation for understanding the universal effects of scaling down of ferroelectrics and will aid the practical realization of nanoscale functional elements. 

\section*{Acknowledgment}

This work was supported by the U.S. Department of Energy, Office of Basic Energy Sciences, Division of Materials Sciences and Engineering under Grant No. DE-SC0005245.  Computational support was provided by the National Energy Research Scientific Computing Center (NERSC), a U.S. Department of Energy, Office of Science User Facility located at Lawrence Berkeley National Laboratory, operated under Contract No. DE-AC02-05CH11231 using NERSC award BES-ERCAP-0025236.x

\newpage

\newpage

\onecolumngrid

\begin{center}
   \textbf{\Large Supplementary Material}
\end{center}

 \renewcommand{\thefigure}{S\arabic{figure}}

\setcounter{figure}{0}

\begin{figure}[h]
\centering
\includegraphics[width=1\textwidth]{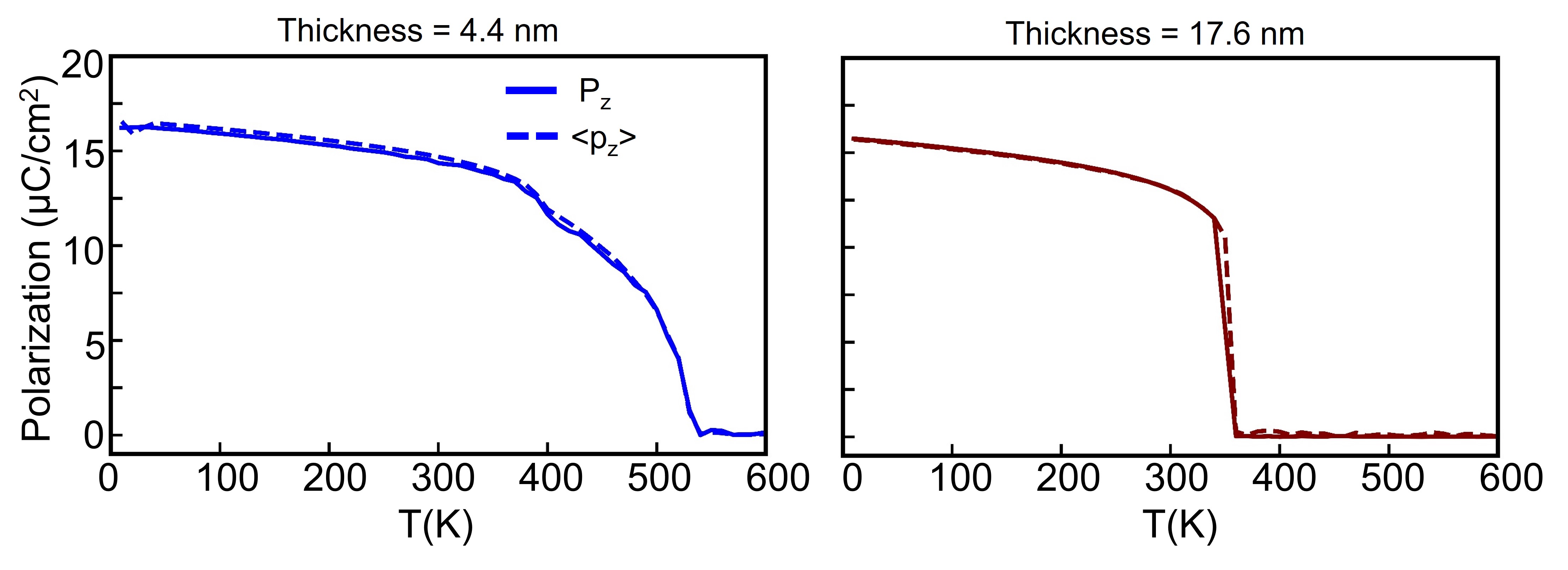}
\caption{ Temperature evolution of z-component of the polarization and local polarization  for two films.  }
\label{figS1}
\end{figure}

\begin{figure}[h]
\centering
\includegraphics[width=1\textwidth]{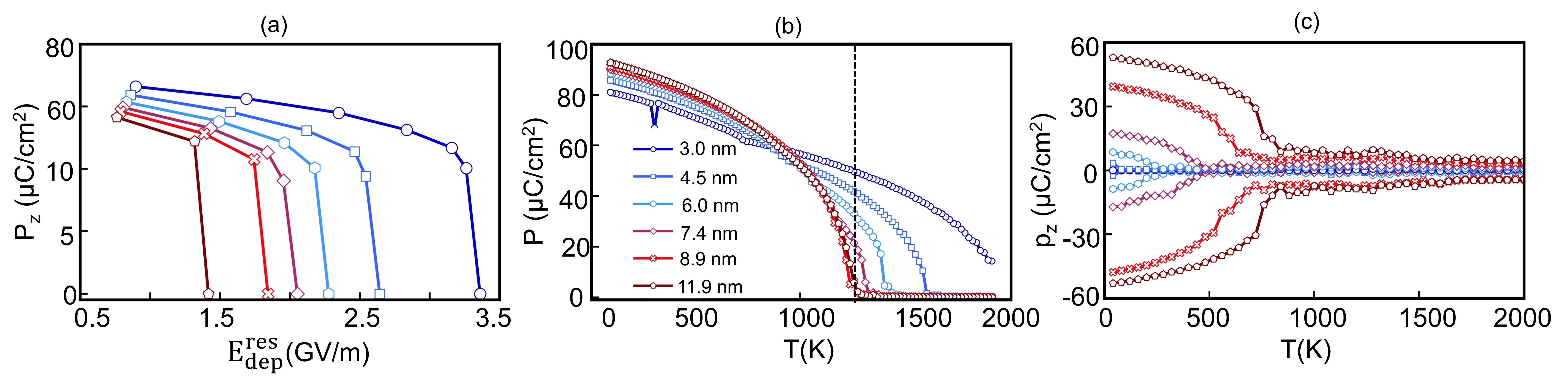}
\caption{ (a) Variation of the z component of the polarization as a function of residual depolarizing field at 10 K for BiFeO$_3$ films. (b) BiFeO$_3$ polarization variation as a function of temperature. (c) The temperature evolution of the local polarization of BiFeO$_3$ films in the nanodomain phase.  }
\label{figS2}
\end{figure}

\begin{figure}
\centering
\includegraphics[width=0.8\textwidth]{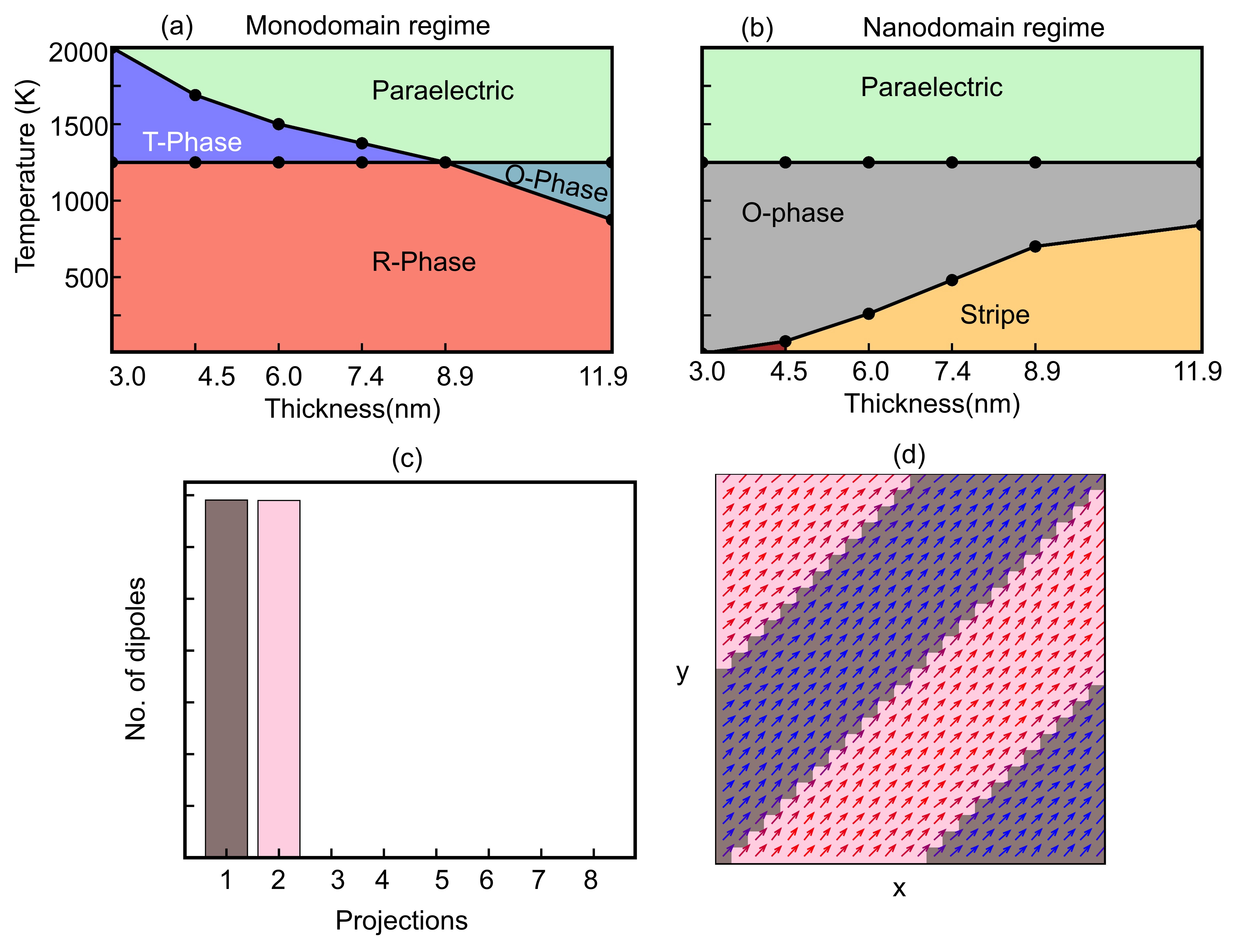}
\caption{ (a)-(b) Temperature-thickness phase diagram  for unstressed BiFeO$_3$ thin films for monodomain and nanodomain phases. (c)-(d) Class distribution of dipole orientations and corresponding domain structures of  films having thickness 8.9 nm.  }
\label{figS3}
\end{figure}

\begin{figure}
\centering
\includegraphics[width=0.8\textwidth]{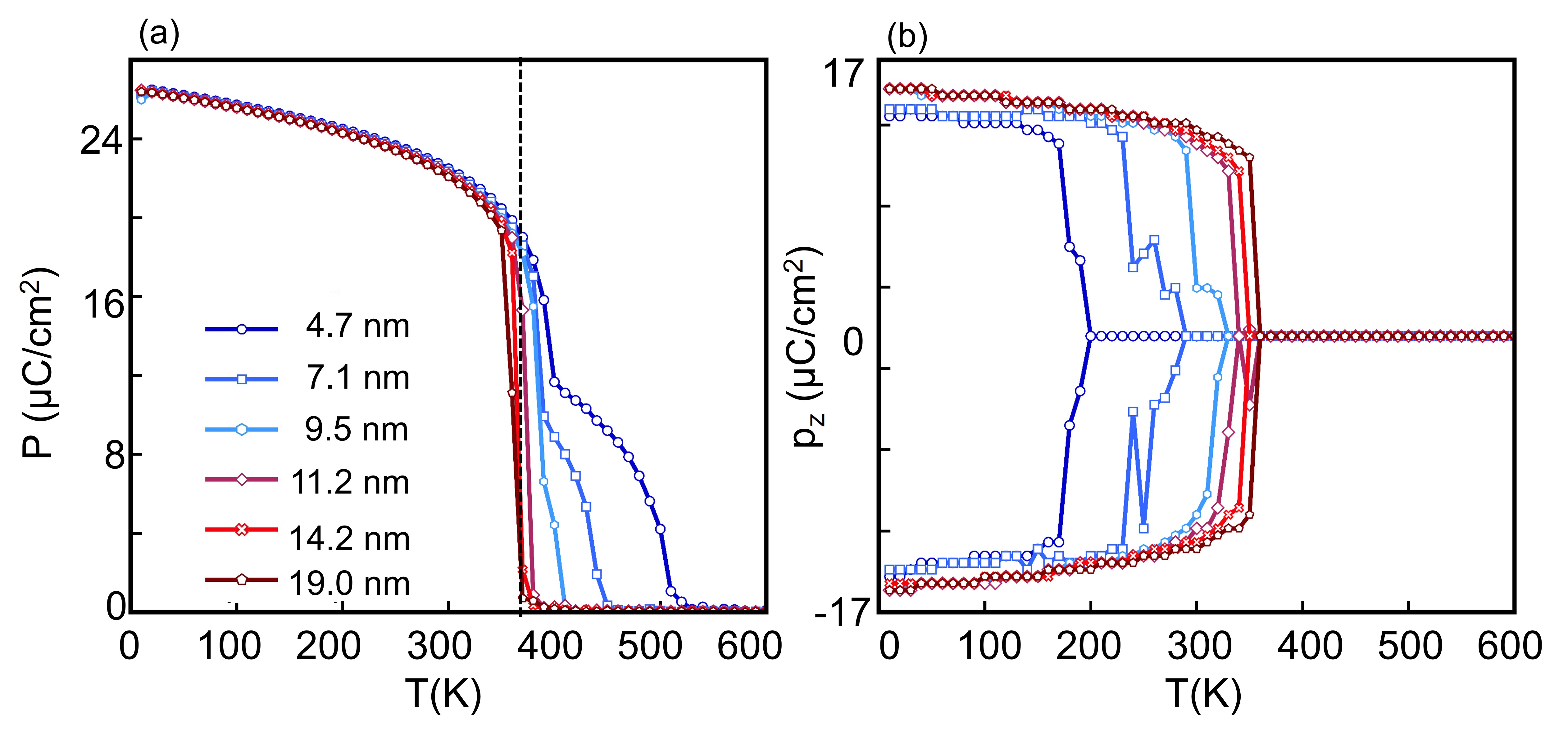}
\caption{(a) CsGeI$_3$ polarization variation as a function of temperature. (b) The temperature evolution of the local polarization of CsGeI$_3$ films in the nanodomain phase. }
\label{figS4}
\end{figure}

\begin{figure}
\centering
\includegraphics[width=0.8\textwidth]{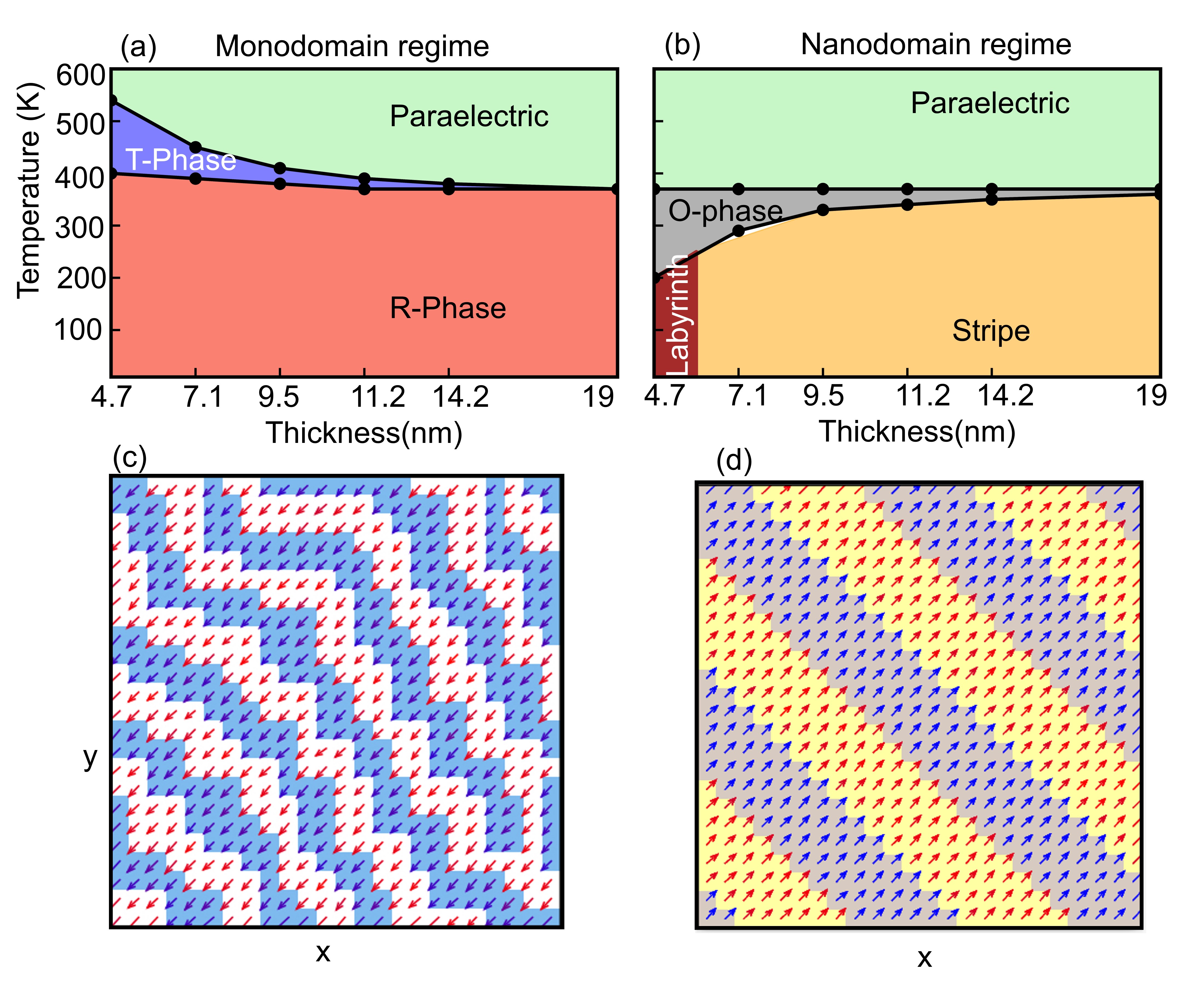}
\caption{(a)-(b) Temperature-thickness phase diagram  for unstressed CsGeI$_3$ thin films for monodomain and nanodomain phases. (c)-(d) Domain structures of films thickness 14.2 nm.     }
\label{figS5}
\end{figure}


\begin{thebibliography}{43}%
\makeatletter
\providecommand \@ifxundefined [1]{%
 \@ifx{#1\undefined}
}%
\providecommand \@ifnum [1]{%
 \ifnum #1\expandafter \@firstoftwo
 \else \expandafter \@secondoftwo
 \fi
}%
\providecommand \@ifx [1]{%
 \ifx #1\expandafter \@firstoftwo
 \else \expandafter \@secondoftwo
 \fi
}%
\providecommand \natexlab [1]{#1}%
\providecommand \enquote  [1]{``#1''}%
\providecommand \bibnamefont  [1]{#1}%
\providecommand \bibfnamefont [1]{#1}%
\providecommand \citenamefont [1]{#1}%
\providecommand \href@noop [0]{\@secondoftwo}%
\providecommand \href [0]{\begingroup \@sanitize@url \@href}%
\providecommand \@href[1]{\@@startlink{#1}\@@href}%
\providecommand \@@href[1]{\endgroup#1\@@endlink}%
\providecommand \@sanitize@url [0]{\catcode `\\12\catcode `\$12\catcode `\&12\catcode `\#12\catcode `\^12\catcode `\_12\catcode `\%12\relax}%
\providecommand \@@startlink[1]{}%
\providecommand \@@endlink[0]{}%
\providecommand \url  [0]{\begingroup\@sanitize@url \@url }%
\providecommand \@url [1]{\endgroup\@href {#1}{\urlprefix }}%
\providecommand \urlprefix  [0]{URL }%
\providecommand \Eprint [0]{\href }%
\providecommand \doibase [0]{http://dx.doi.org/}%
\providecommand \selectlanguage [0]{\@gobble}%
\providecommand \bibinfo  [0]{\@secondoftwo}%
\providecommand \bibfield  [0]{\@secondoftwo}%
\providecommand \translation [1]{[#1]}%
\providecommand \BibitemOpen [0]{}%
\providecommand \bibitemStop [0]{}%
\providecommand \bibitemNoStop [0]{.\EOS\space}%
\providecommand \EOS [0]{\spacefactor3000\relax}%
\providecommand \BibitemShut  [1]{\csname bibitem#1\endcsname}%
\let\auto@bib@innerbib\@empty
\bibitem [{\citenamefont {Scott}(2007)}]{Scott-Apll}%
  \BibitemOpen
  \bibfield  {author} {\bibinfo {author} {\bibfnamefont {J.~F.}\ \bibnamefont {Scott}},\ }\href {\doibase 10.1126/science.1129564} {\bibfield  {journal} {\bibinfo  {journal} {Science}\ }\textbf {\bibinfo {volume} {315}},\ \bibinfo {pages} {954} (\bibinfo {year} {2007})}\BibitemShut {NoStop}%
\bibitem [{\citenamefont {Dawber}\ \emph {et~al.}(2005)\citenamefont {Dawber}, \citenamefont {Rabe},\ and\ \citenamefont {Scott}}]{Scott}%
  \BibitemOpen
  \bibfield  {author} {\bibinfo {author} {\bibfnamefont {M.}~\bibnamefont {Dawber}}, \bibinfo {author} {\bibfnamefont {K.~M.}\ \bibnamefont {Rabe}}, \ and\ \bibinfo {author} {\bibfnamefont {J.~F.}\ \bibnamefont {Scott}},\ }\href {\doibase 10.1103/RevModPhys.77.1083} {\bibfield  {journal} {\bibinfo  {journal} {Rev. Mod. Phys.}\ }\textbf {\bibinfo {volume} {77}},\ \bibinfo {pages} {1083} (\bibinfo {year} {2005})}\BibitemShut {NoStop}%
\bibitem [{\citenamefont {Martin}\ and\ \citenamefont {Rappe}(2016)}]{martin2016thin}%
  \BibitemOpen
  \bibfield  {author} {\bibinfo {author} {\bibfnamefont {L.~W.}\ \bibnamefont {Martin}}\ and\ \bibinfo {author} {\bibfnamefont {A.~M.}\ \bibnamefont {Rappe}},\ }\href@noop {} {\bibfield  {journal} {\bibinfo  {journal} {Nature Reviews Materials}\ }\textbf {\bibinfo {volume} {2}},\ \bibinfo {pages} {1} (\bibinfo {year} {2016})}\BibitemShut {NoStop}%
\bibitem [{\citenamefont {Ahn}\ \emph {et~al.}(2004)\citenamefont {Ahn}, \citenamefont {Rabe},\ and\ \citenamefont {Triscone}}]{science_Rabe}%
  \BibitemOpen
  \bibfield  {author} {\bibinfo {author} {\bibfnamefont {C.~H.}\ \bibnamefont {Ahn}}, \bibinfo {author} {\bibfnamefont {K.~M.}\ \bibnamefont {Rabe}}, \ and\ \bibinfo {author} {\bibfnamefont {J.-M.}\ \bibnamefont {Triscone}},\ }\href {\doibase 10.1126/science.1092508} {\bibfield  {journal} {\bibinfo  {journal} {Science}\ }\textbf {\bibinfo {volume} {303}},\ \bibinfo {pages} {488} (\bibinfo {year} {2004})}\BibitemShut {NoStop}%
\bibitem [{\citenamefont {Rabe}(2005)}]{RABE2005122}%
  \BibitemOpen
  \bibfield  {author} {\bibinfo {author} {\bibfnamefont {K.~M.}\ \bibnamefont {Rabe}},\ }\href {\doibase https://doi.org/10.1016/j.cossms.2006.06.003} {\bibfield  {journal} {\bibinfo  {journal} {Current Opinion in Solid State and Materials Science}\ }\textbf {\bibinfo {volume} {9}},\ \bibinfo {pages} {122} (\bibinfo {year} {2005})}\BibitemShut {NoStop}%
\bibitem [{\citenamefont {Choi}\ \emph {et~al.}(2004)\citenamefont {Choi}, \citenamefont {Biegalski}, \citenamefont {Li}, \citenamefont {Sharan}, \citenamefont {Schubert}, \citenamefont {Uecker}, \citenamefont {Reiche}, \citenamefont {Chen}, \citenamefont {Pan}, \citenamefont {Gopalan}, \citenamefont {Chen}, \citenamefont {Schlom},\ and\ \citenamefont {Eom}}]{Thin_BTO}%
  \BibitemOpen
  \bibfield  {author} {\bibinfo {author} {\bibfnamefont {K.~J.}\ \bibnamefont {Choi}}, \bibinfo {author} {\bibfnamefont {M.}~\bibnamefont {Biegalski}}, \bibinfo {author} {\bibfnamefont {Y.~L.}\ \bibnamefont {Li}}, \bibinfo {author} {\bibfnamefont {A.}~\bibnamefont {Sharan}}, \bibinfo {author} {\bibfnamefont {J.}~\bibnamefont {Schubert}}, \bibinfo {author} {\bibfnamefont {R.}~\bibnamefont {Uecker}}, \bibinfo {author} {\bibfnamefont {P.}~\bibnamefont {Reiche}}, \bibinfo {author} {\bibfnamefont {Y.~B.}\ \bibnamefont {Chen}}, \bibinfo {author} {\bibfnamefont {X.~Q.}\ \bibnamefont {Pan}}, \bibinfo {author} {\bibfnamefont {V.}~\bibnamefont {Gopalan}}, \bibinfo {author} {\bibfnamefont {L.-Q.}\ \bibnamefont {Chen}}, \bibinfo {author} {\bibfnamefont {D.~G.}\ \bibnamefont {Schlom}}, \ and\ \bibinfo {author} {\bibfnamefont {C.~B.}\ \bibnamefont {Eom}},\ }\href {\doibase 10.1126/science.1103218} {\bibfield  {journal} {\bibinfo  {journal} {Science}\ }\textbf {\bibinfo {volume} {306}},\ \bibinfo {pages} {1005} (\bibinfo
  {year} {2004})}\BibitemShut {NoStop}%
\bibitem [{\citenamefont {Ponomareva}\ and\ \citenamefont {Bellaiche}(2006)}]{Growth}%
  \BibitemOpen
  \bibfield  {author} {\bibinfo {author} {\bibfnamefont {I.}~\bibnamefont {Ponomareva}}\ and\ \bibinfo {author} {\bibfnamefont {L.}~\bibnamefont {Bellaiche}},\ }\href {\doibase 10.1103/PhysRevB.74.064102} {\bibfield  {journal} {\bibinfo  {journal} {Phys. Rev. B}\ }\textbf {\bibinfo {volume} {74}},\ \bibinfo {pages} {064102} (\bibinfo {year} {2006})}\BibitemShut {NoStop}%
\bibitem [{\citenamefont {Schlom}\ \emph {et~al.}(2007)\citenamefont {Schlom}, \citenamefont {Chen}, \citenamefont {Eom}, \citenamefont {Rabe}, \citenamefont {Streiffer},\ and\ \citenamefont {Triscone}}]{strain}%
  \BibitemOpen
  \bibfield  {author} {\bibinfo {author} {\bibfnamefont {D.~G.}\ \bibnamefont {Schlom}}, \bibinfo {author} {\bibfnamefont {L.-Q.}\ \bibnamefont {Chen}}, \bibinfo {author} {\bibfnamefont {C.-B.}\ \bibnamefont {Eom}}, \bibinfo {author} {\bibfnamefont {K.~M.}\ \bibnamefont {Rabe}}, \bibinfo {author} {\bibfnamefont {S.~K.}\ \bibnamefont {Streiffer}}, \ and\ \bibinfo {author} {\bibfnamefont {J.-M.}\ \bibnamefont {Triscone}},\ }\href {\doibase 10.1146/annurev.matsci.37.061206.113016} {\bibfield  {journal} {\bibinfo  {journal} {Annual Review of Materials Research}\ }\textbf {\bibinfo {volume} {37}},\ \bibinfo {pages} {589} (\bibinfo {year} {2007})}\BibitemShut {NoStop}%
\bibitem [{\citenamefont {Mehta}\ \emph {et~al.}(1973)\citenamefont {Mehta}, \citenamefont {Silverman},\ and\ \citenamefont {Jacobs}}]{mehta1973depolarization}%
  \BibitemOpen
  \bibfield  {author} {\bibinfo {author} {\bibfnamefont {R.}~\bibnamefont {Mehta}}, \bibinfo {author} {\bibfnamefont {B.}~\bibnamefont {Silverman}}, \ and\ \bibinfo {author} {\bibfnamefont {J.}~\bibnamefont {Jacobs}},\ }\href@noop {} {\bibfield  {journal} {\bibinfo  {journal} {Journal of Applied Physics}\ }\textbf {\bibinfo {volume} {44}},\ \bibinfo {pages} {3379} (\bibinfo {year} {1973})}\BibitemShut {NoStop}%
\bibitem [{\citenamefont {Lyu}\ \emph {et~al.}(2018)\citenamefont {Lyu}, \citenamefont {Estandía}, \citenamefont {Gazquez}, \citenamefont {Chisholm}, \citenamefont {Fina}, \citenamefont {Dix}, \citenamefont {Fontcuberta},\ and\ \citenamefont {Sánchez}}]{dipole-1}%
  \BibitemOpen
  \bibfield  {author} {\bibinfo {author} {\bibfnamefont {J.}~\bibnamefont {Lyu}}, \bibinfo {author} {\bibfnamefont {S.}~\bibnamefont {Estandía}}, \bibinfo {author} {\bibfnamefont {J.}~\bibnamefont {Gazquez}}, \bibinfo {author} {\bibfnamefont {M.~F.}\ \bibnamefont {Chisholm}}, \bibinfo {author} {\bibfnamefont {I.}~\bibnamefont {Fina}}, \bibinfo {author} {\bibfnamefont {N.}~\bibnamefont {Dix}}, \bibinfo {author} {\bibfnamefont {J.}~\bibnamefont {Fontcuberta}}, \ and\ \bibinfo {author} {\bibfnamefont {F.}~\bibnamefont {Sánchez}},\ }\href {\doibase 10.1021/acsami.8b07778} {\bibfield  {journal} {\bibinfo  {journal} {ACS Applied Materials \& Interfaces}\ }\textbf {\bibinfo {volume} {10}},\ \bibinfo {pages} {25529} (\bibinfo {year} {2018})},\ \bibinfo {note} {pMID: 29985584}\BibitemShut {NoStop}%
\bibitem [{\citenamefont {Morikawa}\ \emph {et~al.}(2024)\citenamefont {Morikawa}, \citenamefont {Kodera}, \citenamefont {Shimizu}, \citenamefont {Ishihama}, \citenamefont {Ehara}, \citenamefont {Sakata},\ and\ \citenamefont {Funakubo}}]{dipole-2}%
  \BibitemOpen
  \bibfield  {author} {\bibinfo {author} {\bibfnamefont {T.}~\bibnamefont {Morikawa}}, \bibinfo {author} {\bibfnamefont {M.}~\bibnamefont {Kodera}}, \bibinfo {author} {\bibfnamefont {T.}~\bibnamefont {Shimizu}}, \bibinfo {author} {\bibfnamefont {K.}~\bibnamefont {Ishihama}}, \bibinfo {author} {\bibfnamefont {Y.}~\bibnamefont {Ehara}}, \bibinfo {author} {\bibfnamefont {O.}~\bibnamefont {Sakata}}, \ and\ \bibinfo {author} {\bibfnamefont {H.}~\bibnamefont {Funakubo}},\ }\href {\doibase 10.1063/5.0180449} {\bibfield  {journal} {\bibinfo  {journal} {Applied Physics Letters}\ }\textbf {\bibinfo {volume} {124}},\ \bibinfo {pages} {032901} (\bibinfo {year} {2024})}\BibitemShut {NoStop}%
\bibitem [{\citenamefont {Junquera}\ \emph {et~al.}(2023)\citenamefont {Junquera}, \citenamefont {Nahas}, \citenamefont {Prokhorenko}, \citenamefont {Bellaiche}, \citenamefont {\'I\~niguez}, \citenamefont {Schlom}, \citenamefont {Chen}, \citenamefont {Salahuddin}, \citenamefont {Muller}, \citenamefont {Martin},\ and\ \citenamefont {Ramesh}}]{Polar_Topo}%
  \BibitemOpen
  \bibfield  {author} {\bibinfo {author} {\bibfnamefont {J.}~\bibnamefont {Junquera}}, \bibinfo {author} {\bibfnamefont {Y.}~\bibnamefont {Nahas}}, \bibinfo {author} {\bibfnamefont {S.}~\bibnamefont {Prokhorenko}}, \bibinfo {author} {\bibfnamefont {L.}~\bibnamefont {Bellaiche}}, \bibinfo {author} {\bibfnamefont {J.}~\bibnamefont {\'I\~niguez}}, \bibinfo {author} {\bibfnamefont {D.~G.}\ \bibnamefont {Schlom}}, \bibinfo {author} {\bibfnamefont {L.-Q.}\ \bibnamefont {Chen}}, \bibinfo {author} {\bibfnamefont {S.}~\bibnamefont {Salahuddin}}, \bibinfo {author} {\bibfnamefont {D.~A.}\ \bibnamefont {Muller}}, \bibinfo {author} {\bibfnamefont {L.~W.}\ \bibnamefont {Martin}}, \ and\ \bibinfo {author} {\bibfnamefont {R.}~\bibnamefont {Ramesh}},\ }\href {\doibase 10.1103/RevModPhys.95.025001} {\bibfield  {journal} {\bibinfo  {journal} {Rev. Mod. Phys.}\ }\textbf {\bibinfo {volume} {95}},\ \bibinfo {pages} {025001} (\bibinfo {year} {2023})}\BibitemShut {NoStop}%
\bibitem [{\citenamefont {Fong}\ \emph {et~al.}(2004)\citenamefont {Fong}, \citenamefont {Stephenson}, \citenamefont {Streiffer}, \citenamefont {Eastman}, \citenamefont {Auciello}, \citenamefont {Fuoss},\ and\ \citenamefont {Thompson}}]{Streifer}%
  \BibitemOpen
  \bibfield  {author} {\bibinfo {author} {\bibfnamefont {D.~D.}\ \bibnamefont {Fong}}, \bibinfo {author} {\bibfnamefont {G.~B.}\ \bibnamefont {Stephenson}}, \bibinfo {author} {\bibfnamefont {S.~K.}\ \bibnamefont {Streiffer}}, \bibinfo {author} {\bibfnamefont {J.~A.}\ \bibnamefont {Eastman}}, \bibinfo {author} {\bibfnamefont {O.}~\bibnamefont {Auciello}}, \bibinfo {author} {\bibfnamefont {P.~H.}\ \bibnamefont {Fuoss}}, \ and\ \bibinfo {author} {\bibfnamefont {C.}~\bibnamefont {Thompson}},\ }\href {\doibase 10.1126/science.1098252} {\bibfield  {journal} {\bibinfo  {journal} {Science}\ }\textbf {\bibinfo {volume} {304}},\ \bibinfo {pages} {1650} (\bibinfo {year} {2004})}\BibitemShut {NoStop}%
\bibitem [{\citenamefont {Takahashi}\ \emph {et~al.}(2008)\citenamefont {Takahashi}, \citenamefont {Dahl}, \citenamefont {Eberg}, \citenamefont {Grepstad},\ and\ \citenamefont {Tybell}}]{PBT_stripe}%
  \BibitemOpen
  \bibfield  {author} {\bibinfo {author} {\bibfnamefont {R.}~\bibnamefont {Takahashi}}, \bibinfo {author} {\bibfnamefont { .}~\bibnamefont {Dahl}}, \bibinfo {author} {\bibfnamefont {E.}~\bibnamefont {Eberg}}, \bibinfo {author} {\bibfnamefont {J.~K.}\ \bibnamefont {Grepstad}}, \ and\ \bibinfo {author} {\bibfnamefont {T.}~\bibnamefont {Tybell}},\ }\href {\doibase 10.1063/1.2978225} {\bibfield  {journal} {\bibinfo  {journal} {Journal of Applied Physics}\ }\textbf {\bibinfo {volume} {104}},\ \bibinfo {pages} {064109} (\bibinfo {year} {2008})}\BibitemShut {NoStop}%
\bibitem [{\citenamefont {Lai}\ \emph {et~al.}(2007{\natexlab{a}})\citenamefont {Lai}, \citenamefont {Ponomareva}, \citenamefont {Kornev}, \citenamefont {Bellaiche},\ and\ \citenamefont {Salamo}}]{BTO_stripe}%
  \BibitemOpen
  \bibfield  {author} {\bibinfo {author} {\bibfnamefont {B.-K.}\ \bibnamefont {Lai}}, \bibinfo {author} {\bibfnamefont {I.}~\bibnamefont {Ponomareva}}, \bibinfo {author} {\bibfnamefont {I.~A.}\ \bibnamefont {Kornev}}, \bibinfo {author} {\bibfnamefont {L.}~\bibnamefont {Bellaiche}}, \ and\ \bibinfo {author} {\bibfnamefont {G.~J.}\ \bibnamefont {Salamo}},\ }\href {\doibase 10.1103/PhysRevB.75.085412} {\bibfield  {journal} {\bibinfo  {journal} {Phys. Rev. B}\ }\textbf {\bibinfo {volume} {75}},\ \bibinfo {pages} {085412} (\bibinfo {year} {2007}{\natexlab{a}})}\BibitemShut {NoStop}%
\bibitem [{\citenamefont {Zhang}\ \emph {et~al.}(2017)\citenamefont {Zhang}, \citenamefont {Xie}, \citenamefont {Liu}, \citenamefont {Prokhorenko}, \citenamefont {Nahas}, \citenamefont {Pan}, \citenamefont {Bellaiche}, \citenamefont {Gruverman},\ and\ \citenamefont {Valanoor}}]{Bubble}%
  \BibitemOpen
  \bibfield  {author} {\bibinfo {author} {\bibfnamefont {Q.}~\bibnamefont {Zhang}}, \bibinfo {author} {\bibfnamefont {L.}~\bibnamefont {Xie}}, \bibinfo {author} {\bibfnamefont {G.}~\bibnamefont {Liu}}, \bibinfo {author} {\bibfnamefont {S.}~\bibnamefont {Prokhorenko}}, \bibinfo {author} {\bibfnamefont {Y.}~\bibnamefont {Nahas}}, \bibinfo {author} {\bibfnamefont {X.}~\bibnamefont {Pan}}, \bibinfo {author} {\bibfnamefont {L.}~\bibnamefont {Bellaiche}}, \bibinfo {author} {\bibfnamefont {A.}~\bibnamefont {Gruverman}}, \ and\ \bibinfo {author} {\bibfnamefont {N.}~\bibnamefont {Valanoor}},\ }\href {\doibase https://doi.org/10.1002/adma.201702375} {\bibfield  {journal} {\bibinfo  {journal} {Advanced Materials}\ }\textbf {\bibinfo {volume} {29}},\ \bibinfo {pages} {1702375} (\bibinfo {year} {2017})}\BibitemShut {NoStop}%
\bibitem [{\citenamefont {Nahas}\ \emph {et~al.}(2020)\citenamefont {Nahas}, \citenamefont {Prokhorenko}, \citenamefont {Zhang}, \citenamefont {Govinden}, \citenamefont {Valanoor},\ and\ \citenamefont {Bellaiche}}]{Nahas2020}%
  \BibitemOpen
  \bibfield  {author} {\bibinfo {author} {\bibfnamefont {Y.}~\bibnamefont {Nahas}}, \bibinfo {author} {\bibfnamefont {S.}~\bibnamefont {Prokhorenko}}, \bibinfo {author} {\bibfnamefont {Q.}~\bibnamefont {Zhang}}, \bibinfo {author} {\bibfnamefont {V.}~\bibnamefont {Govinden}}, \bibinfo {author} {\bibfnamefont {N.}~\bibnamefont {Valanoor}}, \ and\ \bibinfo {author} {\bibfnamefont {L.}~\bibnamefont {Bellaiche}},\ }\href {\doibase 10.1038/s41467-020-19519-w} {\bibfield  {journal} {\bibinfo  {journal} {Nature Communications}\ }\textbf {\bibinfo {volume} {11}},\ \bibinfo {pages} {5779} (\bibinfo {year} {2020})}\BibitemShut {NoStop}%
\bibitem [{\citenamefont {Han}\ \emph {et~al.}(2024)\citenamefont {Han}, \citenamefont {Dong}, \citenamefont {Liu},\ and\ \citenamefont {Nie}}]{Free_Mem-1}%
  \BibitemOpen
  \bibfield  {author} {\bibinfo {author} {\bibfnamefont {L.}~\bibnamefont {Han}}, \bibinfo {author} {\bibfnamefont {G.}~\bibnamefont {Dong}}, \bibinfo {author} {\bibfnamefont {M.}~\bibnamefont {Liu}}, \ and\ \bibinfo {author} {\bibfnamefont {Y.}~\bibnamefont {Nie}},\ }\href {\doibase https://doi.org/10.1002/adfm.202309543} {\bibfield  {journal} {\bibinfo  {journal} {Advanced Functional Materials}\ }\textbf {\bibinfo {volume} {34}},\ \bibinfo {pages} {2309543} (\bibinfo {year} {2024})}\BibitemShut {NoStop}%
\bibitem [{\citenamefont {Xu}\ \emph {et~al.}(2020)\citenamefont {Xu}, \citenamefont {Huang}, \citenamefont {Barnard}, \citenamefont {Hong}, \citenamefont {Singh}, \citenamefont {Wong}, \citenamefont {Jansen}, \citenamefont {Harbola}, \citenamefont {Xiao}, \citenamefont {Wang}, \citenamefont {Crossley}, \citenamefont {Lu}, \citenamefont {Liu},\ and\ \citenamefont {Hwang}}]{Xu2020}%
  \BibitemOpen
  \bibfield  {author} {\bibinfo {author} {\bibfnamefont {R.}~\bibnamefont {Xu}}, \bibinfo {author} {\bibfnamefont {J.}~\bibnamefont {Huang}}, \bibinfo {author} {\bibfnamefont {E.~S.}\ \bibnamefont {Barnard}}, \bibinfo {author} {\bibfnamefont {S.~S.}\ \bibnamefont {Hong}}, \bibinfo {author} {\bibfnamefont {P.}~\bibnamefont {Singh}}, \bibinfo {author} {\bibfnamefont {E.~K.}\ \bibnamefont {Wong}}, \bibinfo {author} {\bibfnamefont {T.}~\bibnamefont {Jansen}}, \bibinfo {author} {\bibfnamefont {V.}~\bibnamefont {Harbola}}, \bibinfo {author} {\bibfnamefont {J.}~\bibnamefont {Xiao}}, \bibinfo {author} {\bibfnamefont {B.~Y.}\ \bibnamefont {Wang}}, \bibinfo {author} {\bibfnamefont {S.}~\bibnamefont {Crossley}}, \bibinfo {author} {\bibfnamefont {D.}~\bibnamefont {Lu}}, \bibinfo {author} {\bibfnamefont {S.}~\bibnamefont {Liu}}, \ and\ \bibinfo {author} {\bibfnamefont {H.~Y.}\ \bibnamefont {Hwang}},\ }\href {\doibase 10.1038/s41467-020-16912-3} {\bibfield  {journal} {\bibinfo  {journal} {Nature Communications}\ }\textbf
  {\bibinfo {volume} {11}},\ \bibinfo {pages} {3141} (\bibinfo {year} {2020})}\BibitemShut {NoStop}%
\bibitem [{\citenamefont {Gao}\ \emph {et~al.}(2015)\citenamefont {Gao}, \citenamefont {Chang}, \citenamefont {Ma}, \citenamefont {You}, \citenamefont {Yin}, \citenamefont {Liu}, \citenamefont {Liu}, \citenamefont {Wang},\ and\ \citenamefont {Yuan}}]{gao2015flexible}%
  \BibitemOpen
  \bibfield  {author} {\bibinfo {author} {\bibfnamefont {W.}~\bibnamefont {Gao}}, \bibinfo {author} {\bibfnamefont {L.}~\bibnamefont {Chang}}, \bibinfo {author} {\bibfnamefont {H.}~\bibnamefont {Ma}}, \bibinfo {author} {\bibfnamefont {L.}~\bibnamefont {You}}, \bibinfo {author} {\bibfnamefont {J.}~\bibnamefont {Yin}}, \bibinfo {author} {\bibfnamefont {J.}~\bibnamefont {Liu}}, \bibinfo {author} {\bibfnamefont {Z.}~\bibnamefont {Liu}}, \bibinfo {author} {\bibfnamefont {J.}~\bibnamefont {Wang}}, \ and\ \bibinfo {author} {\bibfnamefont {G.}~\bibnamefont {Yuan}},\ }\href@noop {} {\bibfield  {journal} {\bibinfo  {journal} {NPG Asia Materials}\ }\textbf {\bibinfo {volume} {7}},\ \bibinfo {pages} {e189} (\bibinfo {year} {2015})}\BibitemShut {NoStop}%
\bibitem [{\citenamefont {Zheng}\ \emph {et~al.}(2023{\natexlab{a}})\citenamefont {Zheng}, \citenamefont {Wang}, \citenamefont {Zhang}, \citenamefont {Chen}, \citenamefont {Suo}, \citenamefont {Xing}, \citenamefont {Wang}, \citenamefont {Wei}, \citenamefont {Chen}, \citenamefont {Guo},\ and\ \citenamefont {Wang}}]{halide-rev}%
  \BibitemOpen
  \bibfield  {author} {\bibinfo {author} {\bibfnamefont {W.}~\bibnamefont {Zheng}}, \bibinfo {author} {\bibfnamefont {X.}~\bibnamefont {Wang}}, \bibinfo {author} {\bibfnamefont {X.}~\bibnamefont {Zhang}}, \bibinfo {author} {\bibfnamefont {B.}~\bibnamefont {Chen}}, \bibinfo {author} {\bibfnamefont {H.}~\bibnamefont {Suo}}, \bibinfo {author} {\bibfnamefont {Z.}~\bibnamefont {Xing}}, \bibinfo {author} {\bibfnamefont {Y.}~\bibnamefont {Wang}}, \bibinfo {author} {\bibfnamefont {H.-L.}\ \bibnamefont {Wei}}, \bibinfo {author} {\bibfnamefont {J.}~\bibnamefont {Chen}}, \bibinfo {author} {\bibfnamefont {Y.}~\bibnamefont {Guo}}, \ and\ \bibinfo {author} {\bibfnamefont {F.}~\bibnamefont {Wang}},\ }\href {\doibase https://doi.org/10.1002/adma.202205410} {\bibfield  {journal} {\bibinfo  {journal} {Advanced Materials}\ }\textbf {\bibinfo {volume} {35}},\ \bibinfo {pages} {2205410} (\bibinfo {year} {2023}{\natexlab{a}})}\BibitemShut {NoStop}%
\bibitem [{\citenamefont {Shahrokhi}\ \emph {et~al.}(2020)\citenamefont {Shahrokhi}, \citenamefont {Gao}, \citenamefont {Wang}, \citenamefont {Anandan}, \citenamefont {Rahaman}, \citenamefont {Singh}, \citenamefont {Wang}, \citenamefont {Cazorla}, \citenamefont {Yuan}, \citenamefont {Liu} \emph {et~al.}}]{shahrokhi2020emergence}%
  \BibitemOpen
  \bibfield  {author} {\bibinfo {author} {\bibfnamefont {S.}~\bibnamefont {Shahrokhi}}, \bibinfo {author} {\bibfnamefont {W.}~\bibnamefont {Gao}}, \bibinfo {author} {\bibfnamefont {Y.}~\bibnamefont {Wang}}, \bibinfo {author} {\bibfnamefont {P.~R.}\ \bibnamefont {Anandan}}, \bibinfo {author} {\bibfnamefont {M.~Z.}\ \bibnamefont {Rahaman}}, \bibinfo {author} {\bibfnamefont {S.}~\bibnamefont {Singh}}, \bibinfo {author} {\bibfnamefont {D.}~\bibnamefont {Wang}}, \bibinfo {author} {\bibfnamefont {C.}~\bibnamefont {Cazorla}}, \bibinfo {author} {\bibfnamefont {G.}~\bibnamefont {Yuan}}, \bibinfo {author} {\bibfnamefont {J.-M.}\ \bibnamefont {Liu}},  \emph {et~al.},\ }\href@noop {} {\bibfield  {journal} {\bibinfo  {journal} {Small Methods}\ }\textbf {\bibinfo {volume} {4}},\ \bibinfo {pages} {2000149} (\bibinfo {year} {2020})}\BibitemShut {NoStop}%
\bibitem [{\citenamefont {Zheng}\ \emph {et~al.}(2023{\natexlab{b}})\citenamefont {Zheng}, \citenamefont {Wang}, \citenamefont {Zhang}, \citenamefont {Chen}, \citenamefont {Suo}, \citenamefont {Xing}, \citenamefont {Wang}, \citenamefont {Wei}, \citenamefont {Chen}, \citenamefont {Guo} \emph {et~al.}}]{zheng2023emerging}%
  \BibitemOpen
  \bibfield  {author} {\bibinfo {author} {\bibfnamefont {W.}~\bibnamefont {Zheng}}, \bibinfo {author} {\bibfnamefont {X.}~\bibnamefont {Wang}}, \bibinfo {author} {\bibfnamefont {X.}~\bibnamefont {Zhang}}, \bibinfo {author} {\bibfnamefont {B.}~\bibnamefont {Chen}}, \bibinfo {author} {\bibfnamefont {H.}~\bibnamefont {Suo}}, \bibinfo {author} {\bibfnamefont {Z.}~\bibnamefont {Xing}}, \bibinfo {author} {\bibfnamefont {Y.}~\bibnamefont {Wang}}, \bibinfo {author} {\bibfnamefont {H.-L.}\ \bibnamefont {Wei}}, \bibinfo {author} {\bibfnamefont {J.}~\bibnamefont {Chen}}, \bibinfo {author} {\bibfnamefont {Y.}~\bibnamefont {Guo}},  \emph {et~al.},\ }\href@noop {} {\bibfield  {journal} {\bibinfo  {journal} {Advanced Materials}\ }\textbf {\bibinfo {volume} {35}},\ \bibinfo {pages} {2205410} (\bibinfo {year} {2023}{\natexlab{b}})}\BibitemShut {NoStop}%
\bibitem [{\citenamefont {Zhang}\ \emph {et~al.}(2022)\citenamefont {Zhang}, \citenamefont {Parsonnet}, \citenamefont {Fernandez}, \citenamefont {Griffin}, \citenamefont {Huyan}, \citenamefont {Lin}, \citenamefont {Lei}, \citenamefont {Jin}, \citenamefont {Barnard}, \citenamefont {Raja}, \citenamefont {Behera}, \citenamefont {Pan}, \citenamefont {Ramesh},\ and\ \citenamefont {Yang}}]{CGX_Ferro}%
  \BibitemOpen
  \bibfield  {author} {\bibinfo {author} {\bibfnamefont {Y.}~\bibnamefont {Zhang}}, \bibinfo {author} {\bibfnamefont {E.}~\bibnamefont {Parsonnet}}, \bibinfo {author} {\bibfnamefont {A.}~\bibnamefont {Fernandez}}, \bibinfo {author} {\bibfnamefont {S.~M.}\ \bibnamefont {Griffin}}, \bibinfo {author} {\bibfnamefont {H.}~\bibnamefont {Huyan}}, \bibinfo {author} {\bibfnamefont {C.-K.}\ \bibnamefont {Lin}}, \bibinfo {author} {\bibfnamefont {T.}~\bibnamefont {Lei}}, \bibinfo {author} {\bibfnamefont {J.}~\bibnamefont {Jin}}, \bibinfo {author} {\bibfnamefont {E.~S.}\ \bibnamefont {Barnard}}, \bibinfo {author} {\bibfnamefont {A.}~\bibnamefont {Raja}}, \bibinfo {author} {\bibfnamefont {P.}~\bibnamefont {Behera}}, \bibinfo {author} {\bibfnamefont {X.}~\bibnamefont {Pan}}, \bibinfo {author} {\bibfnamefont {R.}~\bibnamefont {Ramesh}}, \ and\ \bibinfo {author} {\bibfnamefont {P.}~\bibnamefont {Yang}},\ }\href {\doibase 10.1126/sciadv.abj5881} {\bibfield  {journal} {\bibinfo  {journal} {Science Advances}\ }\textbf {\bibinfo
  {volume} {8}},\ \bibinfo {pages} {eabj5881} (\bibinfo {year} {2022})}\BibitemShut {NoStop}%
\bibitem [{\citenamefont {Krishnamoorthy}\ \emph {et~al.}(2015)\citenamefont {Krishnamoorthy}, \citenamefont {Ding}, \citenamefont {Yan}, \citenamefont {Leong}, \citenamefont {Baikie}, \citenamefont {Zhang}, \citenamefont {Sherburne}, \citenamefont {Li}, \citenamefont {Asta}, \citenamefont {Mathews},\ and\ \citenamefont {Mhaisalkar}}]{CGI_solarcell}%
  \BibitemOpen
  \bibfield  {author} {\bibinfo {author} {\bibfnamefont {T.}~\bibnamefont {Krishnamoorthy}}, \bibinfo {author} {\bibfnamefont {H.}~\bibnamefont {Ding}}, \bibinfo {author} {\bibfnamefont {C.}~\bibnamefont {Yan}}, \bibinfo {author} {\bibfnamefont {W.~L.}\ \bibnamefont {Leong}}, \bibinfo {author} {\bibfnamefont {T.}~\bibnamefont {Baikie}}, \bibinfo {author} {\bibfnamefont {Z.}~\bibnamefont {Zhang}}, \bibinfo {author} {\bibfnamefont {M.}~\bibnamefont {Sherburne}}, \bibinfo {author} {\bibfnamefont {S.}~\bibnamefont {Li}}, \bibinfo {author} {\bibfnamefont {M.}~\bibnamefont {Asta}}, \bibinfo {author} {\bibfnamefont {N.}~\bibnamefont {Mathews}}, \ and\ \bibinfo {author} {\bibfnamefont {S.~G.}\ \bibnamefont {Mhaisalkar}},\ }\href {\doibase 10.1039/C5TA05741H} {\bibfield  {journal} {\bibinfo  {journal} {J. Mater. Chem. A}\ }\textbf {\bibinfo {volume} {3}},\ \bibinfo {pages} {23829} (\bibinfo {year} {2015})}\BibitemShut {NoStop}%
\bibitem [{\citenamefont {Lu}\ \emph {et~al.}()\citenamefont {Lu}, \citenamefont {Wang}, \citenamefont {Han}, \citenamefont {Zhao}, \citenamefont {He}, \citenamefont {Song}, \citenamefont {Song},\ and\ \citenamefont {Miao}}]{Rashba_appl}%
  \BibitemOpen
  \bibfield  {author} {\bibinfo {author} {\bibfnamefont {Y.}~\bibnamefont {Lu}}, \bibinfo {author} {\bibfnamefont {Q.}~\bibnamefont {Wang}}, \bibinfo {author} {\bibfnamefont {L.}~\bibnamefont {Han}}, \bibinfo {author} {\bibfnamefont {Y.}~\bibnamefont {Zhao}}, \bibinfo {author} {\bibfnamefont {Z.}~\bibnamefont {He}}, \bibinfo {author} {\bibfnamefont {W.}~\bibnamefont {Song}}, \bibinfo {author} {\bibfnamefont {C.}~\bibnamefont {Song}}, \ and\ \bibinfo {author} {\bibfnamefont {Z.}~\bibnamefont {Miao}},\ }\href {\doibase https://doi.org/10.1002/adfm.202314427} {\bibfield  {journal} {\bibinfo  {journal} {Advanced Functional Materials}\ }\textbf {\bibinfo {volume} {n/a}},\ \bibinfo {pages} {2314427}}\BibitemShut {NoStop}%
\bibitem [{\citenamefont {Kashikar}\ \emph {et~al.}(2024)\citenamefont {Kashikar}, \citenamefont {Lisenkov},\ and\ \citenamefont {Ponomareva}}]{RAVI_CGB}%
  \BibitemOpen
  \bibfield  {author} {\bibinfo {author} {\bibfnamefont {R.}~\bibnamefont {Kashikar}}, \bibinfo {author} {\bibfnamefont {S.}~\bibnamefont {Lisenkov}}, \ and\ \bibinfo {author} {\bibfnamefont {I.}~\bibnamefont {Ponomareva}},\ }\href {\doibase 10.1103/PhysRevB.109.L020101} {\bibfield  {journal} {\bibinfo  {journal} {Phys. Rev. B}\ }\textbf {\bibinfo {volume} {109}},\ \bibinfo {pages} {L020101} (\bibinfo {year} {2024})}\BibitemShut {NoStop}%
\bibitem [{\citenamefont {Ravi}()}]{unpublishedkeyB}%
  \BibitemOpen
  \bibfield  {author} {\bibinfo {author} {\bibfnamefont {K.}~\bibnamefont {Ravi}},\ }\href@noop {} {\enquote {\bibinfo {title} {Finite temperature properties of csgex$_3$ halide ferroelectrics},}\ }\bibinfo {note} {Unpublished}\BibitemShut {NoStop}%
\bibitem [{\citenamefont {Zhong}\ \emph {et~al.}(1994)\citenamefont {Zhong}, \citenamefont {Vanderbilt},\ and\ \citenamefont {Rabe}}]{Effect_H0}%
  \BibitemOpen
  \bibfield  {author} {\bibinfo {author} {\bibfnamefont {W.}~\bibnamefont {Zhong}}, \bibinfo {author} {\bibfnamefont {D.}~\bibnamefont {Vanderbilt}}, \ and\ \bibinfo {author} {\bibfnamefont {K.~M.}\ \bibnamefont {Rabe}},\ }\href {\doibase 10.1103/PhysRevLett.73.1861} {\bibfield  {journal} {\bibinfo  {journal} {Phys. Rev. Lett.}\ }\textbf {\bibinfo {volume} {73}},\ \bibinfo {pages} {1861} (\bibinfo {year} {1994})}\BibitemShut {NoStop}%
\bibitem [{\citenamefont {Zhong}\ \emph {et~al.}(1995)\citenamefont {Zhong}, \citenamefont {Vanderbilt},\ and\ \citenamefont {Rabe}}]{Effective_H1}%
  \BibitemOpen
  \bibfield  {author} {\bibinfo {author} {\bibfnamefont {W.}~\bibnamefont {Zhong}}, \bibinfo {author} {\bibfnamefont {D.}~\bibnamefont {Vanderbilt}}, \ and\ \bibinfo {author} {\bibfnamefont {K.~M.}\ \bibnamefont {Rabe}},\ }\href {\doibase 10.1103/PhysRevB.52.6301} {\bibfield  {journal} {\bibinfo  {journal} {Phys. Rev. B}\ }\textbf {\bibinfo {volume} {52}},\ \bibinfo {pages} {6301} (\bibinfo {year} {1995})}\BibitemShut {NoStop}%
\bibitem [{\citenamefont {Wang}\ \emph {et~al.}(2012)\citenamefont {Wang}, \citenamefont {Weerasinghe},\ and\ \citenamefont {Bellaiche}}]{BFO}%
  \BibitemOpen
  \bibfield  {author} {\bibinfo {author} {\bibfnamefont {D.}~\bibnamefont {Wang}}, \bibinfo {author} {\bibfnamefont {J.}~\bibnamefont {Weerasinghe}}, \ and\ \bibinfo {author} {\bibfnamefont {L.}~\bibnamefont {Bellaiche}},\ }\href {\doibase 10.1103/PhysRevLett.109.067203} {\bibfield  {journal} {\bibinfo  {journal} {Phys. Rev. Lett.}\ }\textbf {\bibinfo {volume} {109}},\ \bibinfo {pages} {067203} (\bibinfo {year} {2012})}\BibitemShut {NoStop}%
\bibitem [{\citenamefont {Lisenkov}\ and\ \citenamefont {Ponomareva}(2009)}]{BST}%
  \BibitemOpen
  \bibfield  {author} {\bibinfo {author} {\bibfnamefont {S.}~\bibnamefont {Lisenkov}}\ and\ \bibinfo {author} {\bibfnamefont {I.}~\bibnamefont {Ponomareva}},\ }\href {\doibase 10.1103/PhysRevB.80.140102} {\bibfield  {journal} {\bibinfo  {journal} {Phys. Rev. B}\ }\textbf {\bibinfo {volume} {80}},\ \bibinfo {pages} {140102} (\bibinfo {year} {2009})}\BibitemShut {NoStop}%
\bibitem [{\citenamefont {Mentzer}\ \emph {et~al.}(2019)\citenamefont {Mentzer}, \citenamefont {Lisenkov}, \citenamefont {Fthenakis},\ and\ \citenamefont {Ponomareva}}]{BZT}%
  \BibitemOpen
  \bibfield  {author} {\bibinfo {author} {\bibfnamefont {C.}~\bibnamefont {Mentzer}}, \bibinfo {author} {\bibfnamefont {S.}~\bibnamefont {Lisenkov}}, \bibinfo {author} {\bibfnamefont {Z.~G.}\ \bibnamefont {Fthenakis}}, \ and\ \bibinfo {author} {\bibfnamefont {I.}~\bibnamefont {Ponomareva}},\ }\href {\doibase 10.1103/PhysRevB.99.064111} {\bibfield  {journal} {\bibinfo  {journal} {Phys. Rev. B}\ }\textbf {\bibinfo {volume} {99}},\ \bibinfo {pages} {064111} (\bibinfo {year} {2019})}\BibitemShut {NoStop}%
\bibitem [{\citenamefont {Waghmare}\ and\ \citenamefont {Rabe}(1997)}]{PTO_phase}%
  \BibitemOpen
  \bibfield  {author} {\bibinfo {author} {\bibfnamefont {U.~V.}\ \bibnamefont {Waghmare}}\ and\ \bibinfo {author} {\bibfnamefont {K.~M.}\ \bibnamefont {Rabe}},\ }\href {\doibase 10.1103/PhysRevB.55.6161} {\bibfield  {journal} {\bibinfo  {journal} {Phys. Rev. B}\ }\textbf {\bibinfo {volume} {55}},\ \bibinfo {pages} {6161} (\bibinfo {year} {1997})}\BibitemShut {NoStop}%
\bibitem [{\citenamefont {Mani}\ \emph {et~al.}(2015)\citenamefont {Mani}, \citenamefont {Lisenkov},\ and\ \citenamefont {Ponomareva}}]{PZo_phase}%
  \BibitemOpen
  \bibfield  {author} {\bibinfo {author} {\bibfnamefont {B.~K.}\ \bibnamefont {Mani}}, \bibinfo {author} {\bibfnamefont {S.}~\bibnamefont {Lisenkov}}, \ and\ \bibinfo {author} {\bibfnamefont {I.}~\bibnamefont {Ponomareva}},\ }\href {\doibase 10.1103/PhysRevB.91.134112} {\bibfield  {journal} {\bibinfo  {journal} {Phys. Rev. B}\ }\textbf {\bibinfo {volume} {91}},\ \bibinfo {pages} {134112} (\bibinfo {year} {2015})}\BibitemShut {NoStop}%
\bibitem [{\citenamefont {Bellaiche}\ \emph {et~al.}(2000)\citenamefont {Bellaiche}, \citenamefont {Garc\'{\i}a},\ and\ \citenamefont {Vanderbilt}}]{PZT}%
  \BibitemOpen
  \bibfield  {author} {\bibinfo {author} {\bibfnamefont {L.}~\bibnamefont {Bellaiche}}, \bibinfo {author} {\bibfnamefont {A.}~\bibnamefont {Garc\'{\i}a}}, \ and\ \bibinfo {author} {\bibfnamefont {D.}~\bibnamefont {Vanderbilt}},\ }\href {\doibase 10.1103/PhysRevLett.84.5427} {\bibfield  {journal} {\bibinfo  {journal} {Phys. Rev. Lett.}\ }\textbf {\bibinfo {volume} {84}},\ \bibinfo {pages} {5427} (\bibinfo {year} {2000})}\BibitemShut {NoStop}%
\bibitem [{\citenamefont {Townsend}\ \emph {et~al.}(2024)\citenamefont {Townsend}, \citenamefont {Kashikar}, \citenamefont {Lisenkov},\ and\ \citenamefont {Ponomareva}}]{Josh_prb}%
  \BibitemOpen
  \bibfield  {author} {\bibinfo {author} {\bibfnamefont {J.}~\bibnamefont {Townsend}}, \bibinfo {author} {\bibfnamefont {R.}~\bibnamefont {Kashikar}}, \bibinfo {author} {\bibfnamefont {S.}~\bibnamefont {Lisenkov}}, \ and\ \bibinfo {author} {\bibfnamefont {I.}~\bibnamefont {Ponomareva}},\ }\href {\doibase 10.1103/PhysRevB.109.094121} {\bibfield  {journal} {\bibinfo  {journal} {Phys. Rev. B}\ }\textbf {\bibinfo {volume} {109}},\ \bibinfo {pages} {094121} (\bibinfo {year} {2024})}\BibitemShut {NoStop}%
\bibitem [{\citenamefont {Thiele}\ \emph {et~al.}(1987)\citenamefont {Thiele}, \citenamefont {Rotter},\ and\ \citenamefont {Schmidt}}]{Thiele}%
  \BibitemOpen
  \bibfield  {author} {\bibinfo {author} {\bibfnamefont {G.}~\bibnamefont {Thiele}}, \bibinfo {author} {\bibfnamefont {H.~W.}\ \bibnamefont {Rotter}}, \ and\ \bibinfo {author} {\bibfnamefont {K.~D.}\ \bibnamefont {Schmidt}},\ }\href {\doibase https://doi.org/10.1002/zaac.19875450217} {\bibfield  {journal} {\bibinfo  {journal} {Zeitschrift für anorganische und allgemeine Chemie}\ }\textbf {\bibinfo {volume} {545}},\ \bibinfo {pages} {148} (\bibinfo {year} {1987})}\BibitemShut {NoStop}%
\bibitem [{\citenamefont {Rapaport}\ and\ \citenamefont {Rapaport}(2004)}]{rapaport2004art}%
  \BibitemOpen
  \bibfield  {author} {\bibinfo {author} {\bibfnamefont {D.~C.}\ \bibnamefont {Rapaport}}\ and\ \bibinfo {author} {\bibfnamefont {D.~C.~R.}\ \bibnamefont {Rapaport}},\ }\href@noop {} {\emph {\bibinfo {title} {The art of molecular dynamics simulation}}}\ (\bibinfo  {publisher} {Cambridge university press},\ \bibinfo {year} {2004})\BibitemShut {NoStop}%
\bibitem [{unp()}]{unpublished_p}%
  \BibitemOpen
  \href@noop {} {\enquote {\bibinfo {title} {Large electrically and chemically tunable rashba-dresselhaus effects in ferroelectric csgex$_3$ (x=cl, br, i) perovskites},}\ }\bibinfo {note} {Unpublished}\BibitemShut {NoStop}%
\bibitem [{\citenamefont {Lai}\ \emph {et~al.}(2007{\natexlab{b}})\citenamefont {Lai}, \citenamefont {Ponomareva}, \citenamefont {Kornev}, \citenamefont {Bellaiche},\ and\ \citenamefont {Salamo}}]{180_stripe}%
  \BibitemOpen
  \bibfield  {author} {\bibinfo {author} {\bibfnamefont {B.-K.}\ \bibnamefont {Lai}}, \bibinfo {author} {\bibfnamefont {I.}~\bibnamefont {Ponomareva}}, \bibinfo {author} {\bibfnamefont {I.}~\bibnamefont {Kornev}}, \bibinfo {author} {\bibfnamefont {L.}~\bibnamefont {Bellaiche}}, \ and\ \bibinfo {author} {\bibfnamefont {G.}~\bibnamefont {Salamo}},\ }\href {\doibase 10.1063/1.2799252} {\bibfield  {journal} {\bibinfo  {journal} {Applied Physics Letters}\ }\textbf {\bibinfo {volume} {91}},\ \bibinfo {pages} {152909} (\bibinfo {year} {2007}{\natexlab{b}})}\BibitemShut {NoStop}%
\bibitem [{\citenamefont {Prosandeev}\ \emph {et~al.}(2010)\citenamefont {Prosandeev}, \citenamefont {Lisenkov},\ and\ \citenamefont {Bellaiche}}]{BFO_Kittel}%
  \BibitemOpen
  \bibfield  {author} {\bibinfo {author} {\bibfnamefont {S.}~\bibnamefont {Prosandeev}}, \bibinfo {author} {\bibfnamefont {S.}~\bibnamefont {Lisenkov}}, \ and\ \bibinfo {author} {\bibfnamefont {L.}~\bibnamefont {Bellaiche}},\ }\href {\doibase 10.1103/PhysRevLett.105.147603} {\bibfield  {journal} {\bibinfo  {journal} {Phys. Rev. Lett.}\ }\textbf {\bibinfo {volume} {105}},\ \bibinfo {pages} {147603} (\bibinfo {year} {2010})}\BibitemShut {NoStop}%
\bibitem [{\citenamefont {C.-M.~Chang}\ and\ \citenamefont {Ponomareva}(2016)}]{BFO_params}%
  \BibitemOpen
  \bibfield  {author} {\bibinfo {author} {\bibfnamefont {S.~L.}\ \bibnamefont {C.-M.~Chang}, \bibfnamefont {B.~K.~Mani}}\ and\ \bibinfo {author} {\bibfnamefont {I.}~\bibnamefont {Ponomareva}},\ }\href {\doibase 10.1080/00150193.2016.1137470} {\bibfield  {journal} {\bibinfo  {journal} {Ferroelectrics}\ }\textbf {\bibinfo {volume} {494}},\ \bibinfo {pages} {68} (\bibinfo {year} {2016})}\BibitemShut {NoStop}%
\end{thebibliography}
\end{document}